\documentclass[traditabstract]{aa}
\usepackage{graphicx}
\usepackage{txfonts}
\usepackage{longtable}
\usepackage[authoryear]{natbib}
\bibpunct{(}{)}{;}{a}{}{,}

\def\ltsima{$\; \buildrel < \over \sim \;$}
\def\simlt{\lower.5ex\hbox{\ltsima}}
\def\gtsima{$\; \buildrel > \over \sim \;$}
\def\simgt{\lower.5ex\hbox{\gtsima}}

\begin{document}

\title{The Potassium abundance in the globular clusters \\ 
NGC~104, NGC~6752 and NGC~6809}
\author{A. Mucciarelli\inst{1,2}, T. Merle\inst{3}, M. Bellazzini\inst{2}}
\offprints{A. Mucciarelli}
\institute{Dipartimento di Fisica e Astronomia, Universit\`a degli Studi di Bologna, 
               Viale Berti Pichat, 6/2, I-40127 Bologna, Italy;
             \email{alessio.mucciarelli2@unibo.it} 
              \and
              INAF - Osservatorio Astronomico di Bologna,
              Via Ranzani 1, 40127 Bologna, Italy
             \and
              Institut d'Astronomie et d'Astrophysique, Universit\'e Libre de Bruxelles, CP.226, Boulevard 
              du Triomphe, 1050, Brussels, Belgium
              }

     \authorrunning{A. Mucciarelli et al.}
   \titlerunning{Potassium in globular clusters}

   \date{Submitted to A\&A }

\abstract{
We derived Potassium abundances in red-giant-branch stars in the Galactic 
globular clusters NGC~104 (144 stars), NGC~6752 (134 stars), and NGC~6809 (151 stars) 
using high-resolution spectra collected with FLAMES at the ESO - Very Large Telescope. 
In the  samples we consider, we do not find significant intrinsic spreads in [K/Fe], which confirms the previous findings , but which is at variance 
with the cases of the massive clusters NGC~2419 and NGC~2808. 
Additionally, marginally significant [K/Fe]-[O/Fe] anti-correlations are found 
in NGC~104 and NGC~6809, and [K/Fe]-[Na/Fe] correlations are found in NGC~104 and 
NGC~6752. No evidence of [K/Fe]-[Mg/Fe] anti-correlation are found.
The results of our analysis are consistent with a scenario in which the process 
leading to the multi-populations in globular clusters also implies  enrichment in the K abundance,
the amplitude of the associated [K/Fe] 
enhancement becoming measurable only in stars showing the most extreme effects of O and Mg depletion. 
Stars enhanced in [K/Fe] have so far only been found in clusters 
harbouring some Mg-poor stars, while the other globulars, without 
a Mg-poor sub-population, show small or null [K/Fe] spreads.}


\maketitle
%

\section{Introduction}
\label{intro}

The last decade of photometric and spectroscopic observations has significantly changed 
our view of the globular clusters (GCs). 
Considered for a long time as the simplest stellar populations available in the 
Universe, with all stars sharing the same age and chemical composition, GCs have revealed 
large and correlated star-to-star inhomogeneities in the abundance of several light elements, like
C, N, Na, O, Mg, Al, and He \citep[see, e.g.,][for a thorough review]{gratton12}.
These inhomogeneities have been traced both in Galactic \citep{carretta09a,carretta09b} and
extragalactic GCs \citep{m09,larsen14}. Their effects are also observed and recognized  in the color magnitude diagrams
of GCs, once the right combination of colours is adopted \citep[see, e.g.,][and references therein]{piot15,milo12, lardo}. 
The photometric evidence, in particular, strongly suggests the presence of multiple discrete stellar populations in GCs 
\citep[see e.g.,][]{milo2808}.

These chemical anomalies are usually interpreted as being due to self-enrichment 
of a gas polluted by the proton capture processes 
(the high temperature extension of the CNO-cycle) that is ejected at a sufficiently
low velocity to be retained in the shallow potential well of proto-GCs. 
The candidate polluters that have been more exhaustively 
considered in the literature are intermediate-mass asymptotic giant branch (AGB) stars
\citep{dantona02,dantona05}, fast rotating massive stars \citep{decressin07}, 
interacting massive binary stars \citep{demink09}, and/or super-massive stars \citep{denis14,denis15}. 

The precise identification of the polluter stars, the timescale of the 
self-enrichment process, and the actual process leading to the observed present-day status 
of GCs are widely debated and we  still lack a convincing comprehensive model
\citep[see, e.g.,][]{decressin07,dercole08,demink09,bastian13,trenti}. As noted, it is far from straightforward to reconcile the overall observational scenario with a
unique self-enrichment process \citep{bastian15}.

Recently, Potassium has joined the exclusive club of chemical elements involved in the chemical 
anomalies in GCs. 
The first evidence of an intrinsic spread in the K abundance has been provided 
for the massive and remote GC NGC~2419. \citet{m12} analysed a sample of 49 giant stars 
by using DEIMOS@Keck low-resolution spectra and finding a huge distribution of 
the [K/Fe] abundance ratio, ranging from solar values up to +2.0 dex. 
Additionally, NGC~2419 shows a large anti-correlation between [K/Fe] and [Mg/Fe], 
which also spans an unusually large ($\sim$2 dex) range of abundances. These results were confirmed by
a high-resolution spectroscopy study of 13  red-giant-branch (RGB) stars by \citet{cohen12}.

Further evidence of an intrinsic spread in the K abundances has been 
provided by \citet{m15} for the massive cluster NGC~2808, one of the few Galactic GCs hosting 
stars with sub-solar [Mg/Fe] abundances, similar to those observed in NGC~2419.
The intrinsic [K/Fe] spread detected in NGC~2808 is significant, but its amplitude is much smaller than
that found in NGC~2419 ($\sim 0.3$~dex vs. $\sim 2.0$~dex). 
Also, [K/Fe] abundance ratios show significant correlations with [Na/Fe] and [Al/Fe], and 
anti-correlations with [O/Fe] and [Mg/Fe]. In particular, all the stars in NGC~2808 with [Mg/Fe]$<$0.0 dex
(hereafter Mg-deficient stars) show K abundances higher than those measured in stars with 
normal Mg abundances, fully analogous to the case of NGC~2419. 
On this basis, \citet{m15} conclude that the [K/Fe] enhancement detected in the Mg-poor stars of these two 
clusters and the K-Mg anti-correlations are ascribable to the same self-enrichment process responsible 
of the observed chemical anomalies. 

\citet{carretta13} measured the K abundance in a handful of stars in each of seven GCs 
(namely, NGC~6752, NGC~6121, NGC~1904, $\omega$ Centauri, 47 Tucanae, NGC~7099, and NGC~6397).
They did not find any evidence of intrinsic K spread in these clusters. 
Moreover, the mean [K/Fe] and [Fe/H] ratios of the considered clusters appear to lie onto the same 
[K/Fe] vs.[Fe/H] relation defined by field stars. However we emphasize that none of the stars studied 
by \citet{carretta13} has [Mg/Fe]$<$0.0. Hence they have not probed the extreme Mg-deficient population 
that is the main driver of the [K/Fe] vs. [Mg/Fe] anti-correlation in NGC~2419 and NGC~2808. 
We point out that the overall scenario of K abundances in Galactic field stars is not well understood, 
 both from the observational and theoretical point of view \citep{romano10}.

Up until now, the only available theoretical study on the K abundance in GCs has been performed 
within the framework of AGB-driven self-enrichment, with the specific aim of reproducing
the Mg-K anti-correlation observed in NGC~2419 \citep{ventura12}. 
In this basic model, super-AGB stars, that ignite the carbon burning in conditions of partial degeneracy 
until the formation of an O-Ne core,  produce fresh Potassium by  
proton captures on Argon nuclei, qualitatively reproducing the observations of NGC~2419 under certain assumptions, 
in particular on the highly uncertain cross section of the $^{38}$Ar$(p,\gamma)^{39}$K reaction.
A similar conclusion was reached by \citet{iliadis16} who investigate the temperature-density 
conditions that are able to reproduce the chemical anomalies observed in NGC~2419: they rule out 
that low-mass, AGB and massive stars can account for the observed chemical patterns, while 
super-AGB stars could be promising polluter stars (but the stellar model parameters need to be 
fine-tuned to produce the required temperatures).

In this paper, within an observational programme aimed at exploring 
the role of Potassium in the process of self-enrichment of GCs,
we investigate the K abundance and its relations with the light elements 
involved in the multiple populations in three GCs, namely NGC~104, 
NGC~6752, and NGC~6809. 

\section{Clusters selection}
The three target clusters have been selected according to the following criteria: 
\begin{enumerate}
\item GCs already analyzed by \citet{carretta09a,carretta09b}, 
i.e., targeting stars whose membership have been already established and whose atmospheric parameters, 
metallicity, and light-elements abundances have been already accurately and homogeneously estimated;
\item GCs covering a wide metallicity range: we selected a metal-rich GC (NGC~104, [Fe/H]=--0.76 dex), 
a metal-intermediate GC (NGC~6752, [Fe/H]=--1.55 dex), and a metal-poor cluster (NGC~6809, [Fe/H]=--1.93 dex), 
according to the metallicity scale provided by \citet{carretta09c};
\item GCs with wide and clear Na-O anti-correlations, signature of a high efficiency
of the self-enrichment process, albeit less extreme than in NGC~2808 and NGC~2419; 
\item GCs close enough to provide relatively bright (V$<$15.5) RGB stars as unchallenging spectroscopic targets.
\end{enumerate}

\section{Observations}
Spectroscopic observations have been secured with the multi-object 
spectrograph FLAMES \citep{pasquini} mounted at the Very Large Telescope of the European 
Southern Observatory. We used FLAMES in the UVES+GIRAFFE combined mode, which 
enables the simultaneous allocation of 132 GIRAFFE-MEDUSA mid-resolution fibers 
and eight UVES high-resolution fibers over a field of view of $\sim$25 arcmin diameter. 
The employed instrumental configurations are the HR18 GIRAFFE setup, covering from 7648 to 7889 \AA\ 
and with a spectral resolution of 18400, and the Red Arm CD\#4 860 UVES setup, 
with a spectral coverage of 6600--10600 \AA\ and a spectral resolution of 47000. 
The adopted UVES and GIRAFFE setups have been chosen because they sample the 
K~I resonance line at 7699 \AA .

To have a complete set of K, O, Na, Mg, and Al abundances for the cluster stars,
we re-observed the same stars already observed by  \citet{carretta07}, 
\citet{carretta09a}, and \citet{carretta09b} 
in their spectroscopic survey devoted to study Na and O abundances in GCs
(hereafter, we refer to this spectroscopic survey as CAR).
We refer to these papers for the description of the targets selection.
Owing to some FLAMES fibers currently broken or parked, a few stars
previously observed cannot be duplicated in our observations. 
For each cluster, two target configurations have been observed to guarantee 
the maximum overlap with the previous observations. 
Two observations of 1330 s each were secured for each target configuration.

Spectral reduction has been performed with the dedicated ESO 
pipelines\footnote{http://www.eso.org/sci/software/pipelines/}, 
including bias subtraction, flat-fielding, wavelength calibration with a reference 
Th-Ar lamp, extraction of the 1-dimensional spectra and (only for UVES spectra) 
merging of the overlapping orders. 

In a few cases, where the K~I line is damaged for the occurrence of a cosmic ray
in one of the exposures, only the other spectrum was used.
For all the other cases, the individual exposures have been coadded together, 
after which they were corrected for their proper heliocentric velocity. 
The final spectra have on average S/N  per pixel  of 
between $\sim$60-70 for the faintest stars (V$\sim$15.5) and up to $\sim$400 
for the brightest targets (V$\sim$11.5).

We summarize the key information about the observations of the target clusters below. We note that we excluded from our analysis the stars identified by CAR as field stars 
according to their radial velocity or with uncertain/unreliable (V-K) colors, hence 
with uncertain atmospheric parameters.
\begin{enumerate}
\item {\sl NGC~104 --}
A total of 145 giant stars have been observed with GIRAFFE, 
11 of them have also been observed  with UVES. 
Two stars of \citet{carretta09b} are not included in our sample, 
namely \#24553 and \#35454.
\item {\sl NGC~6752 --}
A total of 135 stars were observed with GIRAFFE, seven of them 
also observed  as UVES targets, while  eight other stars were observed 
only with UVES. Only the star \#19714 of \citet{carretta07} was not 
observed. 
\item {\sl NGC~6809 --}
A total of 153 stars were observed with GIRAFFE.
Three stars of the original sample of \citet{carretta09a} 
(namely \#7000296, \#7000057, and \#6000017) did not 
reply.
Thirteen stars were taken with UVES and only one of them is not 
in common with the GIRAFFE sample.
\end{enumerate}

\section{Radial velocities}
Radial velocities were measured with DAOSPEC \citep{stetson} 
using tens of metallic lines available in the considered spectral range.
The accuracy of the wavelength calibration 
has been checked by measuring the position of the emission sky lines 
and comparing them with their rest-frame position listed in \citet{oster96}, 
finding no significant shift.
Typical uncertainties in radial velocities, computed as the dispersion of the mean 
divided by the root mean squares of the number of lines, are $\sim$0.4-0.7 km/s.
The measured radial velocities have been compared with those obtained by CAR, 
to identify possible binary stars. In the distribution of radial velocity
differences between our measures and those by CAR ($\Delta_{RV}$) as a function
of V magnitude, we identified four $>3\sigma$ outliers, which we classify as 
candidate binary systems: one star in NGC~104 (\#22726), one star in NGC~6752 (\#21828), and two stars 
in NGC~6809 (\#7000106 and \#7000492). These stars have been excluded from the following analysis.
The observed differences in radial velocity between the two epochs (in the sense this study - CAR) 
for these stars are $\Delta_{RV}=-14.3$~km/s, $\Delta_{RV}=-7.6$~km/s, $\Delta_{RV}=+5.5$~km/s and  
$\Delta_{RV}=+2.9$~km/s, respectively, where the typical dispersion in $\Delta_{RV}$ is $<1.0$~km/s, 
nearly independent of the star magnitude.

The average differences between our measures and those by CAR are (excluding the candidate 
binary stars) +0.01 km/s ($\sigma$=~0.40 km/s) for NGC~104,
--0.18 km/s ($\sigma$=~0.67 km/s) for NGC~6752 and
--0.15 km/s ($\sigma$=~0.78 km/s) for NGC~6809.
We note that the standard deviation increases as the cluster metallicity decreases, 
owing to the associate decrease in the number 
of spectral lines that can be used to estimate radial velocities.

\section{Chemical analysis}
\label{chems}

The K abundances were derived with the code GALA \citep{m13g}, 
by matching the measured and theoretical equivalent widths (EWs) 
of the resonance K line at 7699 \AA . 
Model atmospheres have been obtained by interpolating in the 
MARCS model grid with standard chemical composition 
\citep{gustaf08} and assuming an overall metallicity [M/H] that 
matches the average iron abundance of each cluster, as provided 
by \citet{carretta09c}. 
The adopted oscillator strength for the K line at 7699 \AA\ is $\log(gf)$=--0.176 from 
the NIST database while, for the Van der Waals damping constant, we used the the value quoted 
by \citet{barklem00}, namely $\log\gamma_{VdW}/N_{H}$=~7.445. 
The adopted solar abundance is K$_{\odot}=$5.11 \citep{caffau11}.
The K abundances of the stars of a given cluster were rescaled 
adopting the average [Fe/H] abundance, as quoted by \citet{carretta09c}.


EWs have been measured with the code DAOSPEC, launched automatically 
with the wrapper 4DAO \citep{m13_4dao} that also enables  a visual inspection 
of the best-fit profile for each individual line.
DAOSPEC performs the line-profile fitting under the assumption of a Gaussian profile.
As a sanity check, we remeasured under IRAF the EW of some strong K~I lines with 
EW larger than $\sim$200 m\AA,\, assuming a Voigt profile. The difference 
in the measured EWs are negligible, of the order of 1 m\AA\ or less, 
confirming that the assumption of a Gaussian profile is adequate for the strongest K lines 
at the used spectral resolution. 

Because the K line at 7699 \AA\ is located on the red side of the telluric A band, 
we carefully checked in all the target stars whether the K line was blended 
with a telluric feature. According to their radial velocity, only some stars of 
NGC~6809 are marginally contaminated. To clean the K line, we used 
appropriate synthetic spectra of the atmospheric transmission calculated with the {\tt TAPAS} tool 
\citep{bertaux14}, taking into account the specific observing conditions of our data.

As summarised above, a total of 18 stars were observed both with UVES and GIRAFFE fibers. 
To check the stability of the measured EWs with respect to the 
spectral resolution (in particular related to the continuum placement),
we compared the EWs of the K line measured with UVES and GIRAFFE. 
An average difference of --0.02$\pm$0.8 m\AA\  ($\sigma$=~4.5 m\AA\  ) is found, 
ensuring that no systematic difference  exist between the two sets of EWs.

Table~1 lists the adopted atmospheric parameters, the measured EWs, and the derived [K/Fe] abundance ratios 
and the corresponding uncertainties for each target star.

\subsection{Atmospheric parameters}
\begin{figure}
\includegraphics[width=84mm]{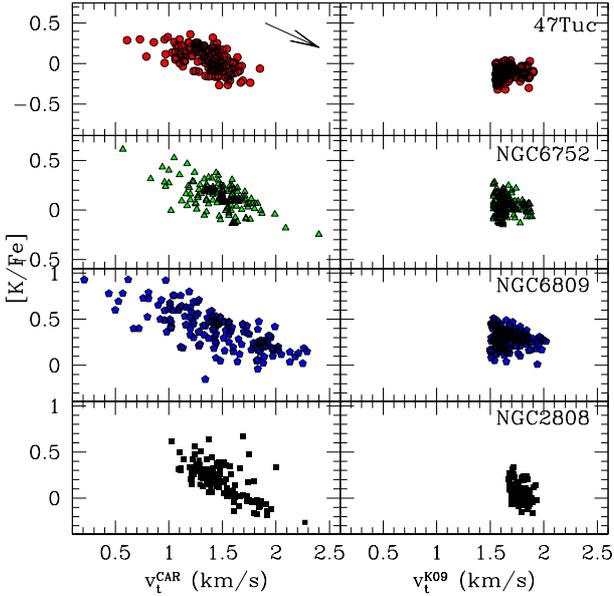}
\caption{Left panels: behaviour of the [K/Fe] as a function of $v_{t}$ provided by CAR. 
Red circles are the stars of NGC~104, green triangles those of NGC~6752, blue pentagons 
those of NGC~6809.
Right panels: behaviour of the [K/Fe] as a function of $v_{t}$ derived according to \citet{kirby09}. 
The stars of NGC~2808 \citep[black squares,][]{m15} are shown for comparison in the lowest panels.
The arrow in the upper-left panel shows the effect on [K/Fe] of the change in $v_{t}$.}
\label{vturb1}
\end{figure}

The adopted effective temperatures ($T_\mathrm{eff}$) and  surface gravities (log~g) are those derived 
photometrically by CAR.
The adoption of the microturbulent velocity ($v_{t}$) for these stars require special care.
The K line is a saturated line (up to $\sim250$ m\AA\ for the brightest stars in 
NGC~104, the most metal-rich GC of the sample) located close or along the flat portion 
of the curve of growth; thus, 
the abundances derived from this transition can be very sensitive to the 
microturbulent velocity (at variance with the weaker lines located on the linear 
part of the curve of growth).
Our first attempt is to use $v_t$ derived by CAR by imposing no trend 
between the theoretical EWs and the abundance of the iron lines. 
Adopting these values, for all the clusters, we found a clear trend between 
the [K/Fe] abundance ratio and $v_t$ (see left panels of Fig.~\ref{vturb1}), 
while no trend is found between [K/Fe] and $T_\mathrm{eff}$ and between [K/Fe] and log~g.  
Also, we note that the slope between [K/Fe] and $v_t$ is fully compatible with the 
variation of [K/Fe] owing to a change of this parameter (the effect on [K/Fe] of a change 
of $v_t$ is shown as an arrow in the upper-left panel of Fig.~\ref{vturb1}).
As explained by \citet{carretta14b} concerning the $v_t$ derived using weak lines, 
"the values of $v_t$  obtained with this 
technique are unsuitable when applied 
to strong lines such as the Na D or the Ba lines, 
resulting in strong trends as a function of the microturbulent velocity". 
In fact, we checked that the EW of the K line is always stronger than 
that of the strongest Fe~I lines used by CAR to infer $v_t$ 
(E. Carretta, private communication).
All these evidence suggest that the $v_t$ scale derived by CAR 
is not appropriate for a saturated line as the K line used in this work 
(but they remain appropriate for weaker transitions).

A similar result has been found by \citet{worley13} 
that determined Ba abundances for giant stars in the globular cluster M15, 
testing different assumptions of $v_t$ (see their Fig.~3). Similar to the K line,  
also the Ba lines are strong and sensitive to $v_t$. \citet{worley13}  
find a trend between Ba abundances and $v_t$ when the original $v_t$ values inferred by CAR are adopted.
The same approach was adopted by \citet{carretta14b}.
In all  three cluster targets the $v_t$ by CAR span large ranges 
of values (more than 1 km/s, with the extreme case of NGC~6809, where the 
stars range from 0.21 to 2.29 km/s). Such wide ranges seems remarkably
unlikely, in particular with respect to the values of $v_t$ obtained from 
high-resolution spectra, hence based on a large number of iron lines 
\citep[see e.g.][]{carretta09b}. Also,
our target stars cover a small region in the $T_\mathrm{eff}$--logg space 
(only the brightest portion of the RGB), hence they are expected to cover 
a smaller range of $v_t$.

\begin{figure}
\includegraphics[width=84mm]{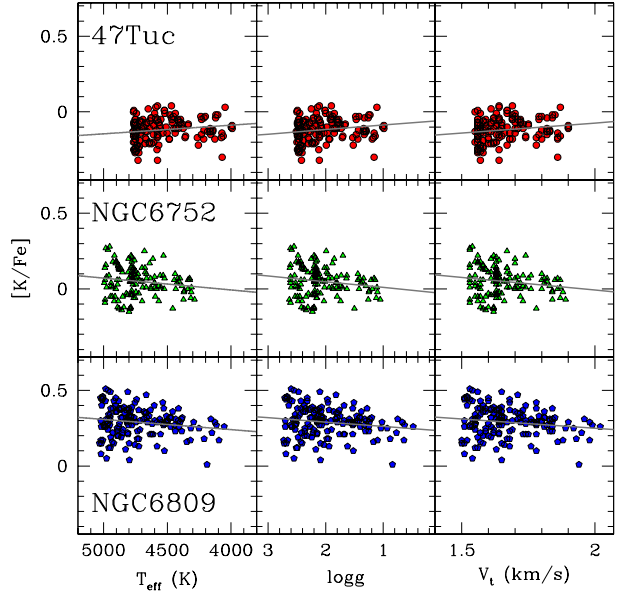}
\caption{Behaviour of the [K/Fe] as a function of  $T_\mathrm{eff}$ (left panels), 
log~g (middle panels), and  $v_{t}$ (right panels). Grey lines represent the linear fits.}
\label{trend_tg}
\end{figure}

As per \citet{m15}, we adopt the relation by \citet{kirby09} that provides 
$v_{t}$ as a function of log~g. 
This relation was obtained with a linear fit between logg and $v_t$ (the latter 
derived spectroscopically) of a compilation of high-resolution measurements available 
in literature (see the references in Kirby et al.2009).
The $v_t$ values obtained with this relation range from $\sim$1.5 km/s to $\sim$2 km/s; 
the cut at $\sim$1.5 km/s visible in all the clusters reflects the similar 
luminosity (and gravity) limit reached by the faintest targets of the three GCs.
The [K/Fe] abundance ratios derived with 
the \citet{kirby09} calibration do not exhibit peculiar or significant 
trend with $T_\mathrm{eff}$, log~g and $v_t$, 
as shown in Fig.~\ref{trend_tg}.

\subsection{Uncertainties}
\label{errors}
Two sources of uncertainties in the K abundance were taken into account, 
one arising from the EW measurement and one arising from the uncertainties 
in the adopted atmospheric parameters. 
The uncertainty in the EW measurement of the K line was estimated 
by DAOSPEC according to the residual of the line fitting and then 
translated in abundance uncertainty. Owing to the high S/N of the spectra and 
the strength of the K line, the abundance uncertainties associated 
with the measurement errors range from $\sim$0.01 to $\sim$0.05 dex.

The K abundance derived from the line at 7699 \AA\ is mainly sensitive
to $T_\mathrm{eff}$ and $v_t$. We adopted the typical uncertainties in $T_\mathrm{eff}$ quoted by CAR 
(40 K for NGC~104, 58 K for NGC~6752, and NGC~6809):
a change in $T_\mathrm{eff}$ of $\pm$50 K changes the K abundance of about $\pm$0.07 dex. 
The typical uncertainties in log~g quoted by CAR (0.06 for all these three clusters) 
have a negligible impact (at a level of 0.01 dex) on the K abundance. On the other hand, 
this type of variation of log~g leads to a change in $v_t$ of $\sim$0.01-0.02 km/s, according to the relation by \citet{kirby09}.
Considering this variation and the uncertainty of the relation by \citet{kirby09}, we estimated 
a typical internal uncertainty in $v_t$ of $\pm$0.05 km/s that changes the K abundance of $\mp$0.03 dex. 
The total uncertainty for [K/Fe] is typically about 0.10-0.12 dex.


\section{NLTE corrections}
\label{nlte}

We consistently solved  the statistical equilibrium and radiative transfer for 
populations and lines of the K~I species using the NLTE radiative transfer code 
MULTI (version 2.3) \citep{carlsson86,carlsson92}.
Since the K~I species is like a trace element, the possible NLTE effects on the atmosphere 
was neglected as is usual. We use LTE model atmospheres from MARCS 
\citep{gustaf08}. Continuous and line opacities for the other species 
are treated in LTE using the Upssala package for the former as included in MULTI, 
and VALD atomic data for the latter.
The K model atom has been computed with the automatic tool {\tt FORMATO} 
(T. Merle et al. in prep.)
For the purpose of the study, we only need to consider the fine structure 
of the first excited level of K to recover properly the resonant doublet. 
Energy levels from NIST were used and completed with levels 6g $^2$G and 
6h $^2$H$^\circ$ from \citet{park71} as done in \citet{bruls92}.
The radiative bound-bound transitions between the mean levels have been obtained from 
the combination of the individual lines from the VALD database. In addition, we add nine multiplets 
(doublets and triplets) from \citet{bruls92} not present in the VALD 
database and some of them using the added levels 6g~$^2$G and 6h~$^2$H$^\circ$. 
For the 7699~\AA\ line, we used the same atomic data used in the chemical analysis (see Section~\ref{chems}).

The bound-bound electron collision transitions were computed
using the Van Regemorter formula \citep{vanregemorter}, based
on the oscillator strength of the corresponding radiative transitions. 
For collision transition without radiative counterpart, a default 
value for the effective collision strength is set to unity. The
bound-free electron collisions were computed with the Seaton
formula \citep{seaton}, using the radiative photoionization cross-section counterpart at the threshold.

The grid of NLTE correction for the K abundances from the 7699~\AA\ line used in this study
was calculated on the basis of the range of atmospheric parameters of the observed GCs. 
In particular, seven effective temperatures $T_\mathrm{eff}$=~3900, 4000, 4250, 4500, 4750, 5000, 5250~K, 
six surface gravities log~g={0.5, 1.0, 1.5, 2.0, 2.5, 3.0~dex, three overall metallicities corresponding 
to the mean metallicity of the GCs, and two microturbulent 
velocities $v_t$=~1 and  2 km/s, were  considered. For each metallicity, we also vary the K abundance 
by step of 0.3~dex, from [K/Fe]=--0.9 and +0.9 dex. 
For each target star,  we calculated the corresponding NLTE abundance correction by interpolating into 
the grid to find the NLTE K abundance that fits the corresponding observed equivalent width of the K I 7699
\AA\ line.

The NLTE abundance corrections for this line ($\Delta
A(\mathrm{K}) = A(\mathrm{K})_\mathrm{NLTE} -A(\mathrm{K})_\mathrm{LTE}$) 
depend strongly on the atmospheric parameters as shown in the first three panels of Fig.~\ref{nltr}. 
The corrections are always negative, meaning that
the K abundance is always overestimated in LTE owing to the scattering in
this resonance line. For the range of the atmospheric parameters
considered for the three GCs, the abundance corrections is 
between $-0.54$ and $-0.03$ dex. The correction increases with
$T_\mathrm{eff}$ and $\log{g}$ and decreases with the K abundance.
We estimate that the typical uncertainties in the atmospheric parameters 
quoted in Section~\ref{errors} translate into variations of $\Delta A(\mathrm{K})$ 
of the order of $\pm$0.02-0.03 dex.

The main sources of uncertainties in the calculation of NLTE corrections 
come from photoionizations and H collisions. 

No quantum mechanical data exist for the photoionization of the energy
levels of this element. The radiative bound-free transition of the
ground stage come from a fit of experimental measurements. At the
photoionisation threshold (2706.36 \AA), \citet{sandner} provides a 
cross-section of $\sigma=3.5\times10-22cm^{-2}$. For the dependance of the 
cross-section away from threshold, we used the values from 
\citep{rahman} based on the quantum-defect method. For all other levels,
we used hydrogenic approximation (Kramer's law) with a bound-free Gaunt 
factor depending on principal and secondary quantum numbers computed
thanks to the formulation of \cite{karzas} implemented by
\citet{janicki}.

The sensitivity to the H collisions has also been investigated. Using
the Drawin formula \citep{drawin} with a scaling factor of 1, we recalculate a
complete grid and re-estimate the complete NLTE abundance corrections.
We show the comparison in the lower-right panel of Fig.~\ref{nltr} between these two sets of
corrections. The NLTE abundance correction is slightly reduced when 
inelastic hydrogen collisions are included.
The average K abundances vary by less than 0.05 dex and
all the results discussed in the next section are not affected by this choice.

Nevertheless, we emphasise that the Drawin's formula
does not have a correct physical background. It has recently been shown by
\citet{barklem11} that the Drawin's formula compares poorly with
the results of available full quantum scattering calculations based on
detailed quantum mechanical studies. But in the absence of any quantum
mechanical data for collisions between potassium and hydrogen, the
Drawin's formula was used to estimate the possible uncertainties
in the NLTE calculations.

\begin{figure}
\includegraphics[width=84mm]{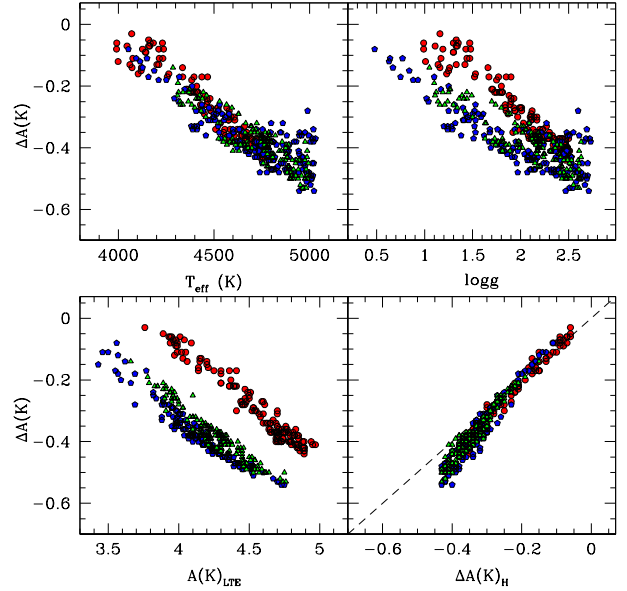}
\caption{Behaviour of the difference between NLTE and LTE Potassium abundance 
(calculated without including H collisions)
as a function of $T_\mathrm{eff}$ (upper-left panel), logg (upper-right panel) and $A(K)_\mathrm{LTE}$ 
(lower-left panel). Lower-right panel shows the comparison between the NLTE correction 
calculated with and without the inclusion of H collisions.
Same symbols used as Fig.~\ref{vturb1}.}
\label{nltr}
\end{figure}


\section{Potassium abundances}
\label{kabs}

\subsection{Potassium abundance spread}

The mean K abundance of each cluster was calculated with the maximum likelihood 
algorithm described in \citet{m12}, which also provides  the intrinsic spread ($\sigma_{int}$) 
and the associated uncertainty, calculated by taking into account the uncertainties for each individual star. 
Table~2 lists the derived average abundance for each cluster, together with the value 
of the observed and intrinsic dispersion, the number of stars and the cluster metallicity.

In the case of NGC~6809, the derived average abundance is $\langle$[K/Fe]$\rangle$=+0.29$\pm$0.01 dex 
with an observed scatter of $\sigma_{obs}$=0.10 dex and an intrinsic 
scatter of $\sigma_{int}$=0.02$\pm$0.02 dex. NGC~6809 is not included in the sample analysed by \citet{carretta13} 
and no additional measures of K abundances are available in the literature.

For NGC~6752 we derived an average abundance of $\langle$[K/Fe]$\rangle$=+0.04$\pm$0.01 dex, with an observed 
dispersion $\sigma_{obs}$=~0.10 dex.  Also for this cluster, the observed scatter is 
compatible with null intrinsic scatter, with $\sigma_{int}$=0.02$\pm$0.03 dex. 
\citet{carretta13} derive an abundance of $\langle$[K/Fe]$\rangle$=+0.17$\pm$0.04 dex from 25 turnoff and RGB stars.

For NGC~104, we derived a mean abundance $\langle$[K/Fe]$\rangle$=--0.12$\pm$0.01 dex, with an observed 
dispersion $\sigma_{obs}$=~0.08 dex that is fully compatible with a zero-scatter 
($\sigma_{int}$=0.00$\pm$0.02 dex). 
The mean derived by \citet{carretta13} ($\langle$[K/Fe]$\rangle$=+0.14$\pm$0.03) from the analysis of 12 turnoff 
and sub-giant stars (no star is in common with our sample) is formally incompatible with our value. 
However, we check that the differences with \citet{carretta13} for NGC~6752 and NGC~104 can be easily ascribed to
differences in the adopted model atmospheres (ATLAS9 vs. MARCS) and NLTE corrections (\citealt{takeda02} vs. our own computations, 
see Sect.~\ref{nlte}).

In conclusion, we confirm  the results by \citet{carretta13}, i.e.,  that these clusters 
do not display a measurable intrinsic scatter in the abundance of Potassium. 
Also, in this case, we note that no Mg-deficient star is known in the GCs analysed here \citep{carretta09b}.


\subsection{Potassium and light element abundances}

Figures~\ref{kna}, \ref{ko}, \ref{kmg}, and \ref{kal} show the 
behaviour of [K/Fe] as a function of  [Na/Fe], [O/Fe], [Mg/Fe] and [Al/Fe] 
(as derived by CAR) for the three clusters considered here plus NGC~2808
\citep[from][]{m15}, taken as a reference case ([Fe/H]=--1.18 dex).
For each 
pair [K/Fe]-[X/Fe] (with X indicating Na, O, Mg and Al) we calculated 
the non-parametric Spearman correlation coefficient ($C_S$) and the corresponding two-tailed probability $P$
that a $C_S$ larger than (or equal to) the observed one (in absolute value) may arise from non-correlated random variables. 
The results are listed in Table~3.

Figures~\ref{kna} and \ref{ko}, where  the large samples enable a deeper insight, deserve some 
comment. 
[K/Fe] is found to correlate with [Na/Fe] in NGC~6752 and NGC~104, while no correlation is found in NGC  6809.
On the other hand, [K/Fe] is found to anti-correlate [O/Fe] in NGC~104, while the other two clusters 
show weak hints of anti-correlation. The behaviour is very similar to that observed in NGC~2808,  
with smaller amplitudes, and, in some cases, the correlation seems marginally significant.  
In particular, the probability that the $C_S$ observed for the [K/Fe] vs. [O/Fe] correlation in NGC~104 occurred by chance 
is as low as 0.35 per cent.

We checked whether the error distribution of the atmospheric parameter with the largest
impact on the abundance estimates ($T_\mathrm{eff}$) can introduce spurious correlations between
the considered abundance ratios via its known correlation with other 
parameters. 
An increase of $T_\mathrm{eff}$ corresponds to an increase of log~g and 
a decrease of $v_t$. These correlated changes of the atmospheric parameters lead to an increase of 
all the abundance ratios that we consider 
in this work. Thus, the uncertainties in the parameters could mimic a positive
[K/Fe]-[Na/Fe] correlation but they are not able to introduce a [K/Fe]-[O/Fe] anti-correlation. 

We remark that the Spearman 
correlation coefficient does not take into account the internal uncertainties 
in the individual abundance, at variance with the estimate of the intrinsic 
spread. This may suggest a possible overestimate of the internal uncertainties on individual [K/Fe] measures, which
 can hide small but real intrinsic spreads. 
For instance, if we decrease the internal uncertainty in $v_t$ by a factor of 2, and repeat the 
maximum likelihood analysis, a non-null [K/Fe] spread of 0.04-0.05~dex 
is found in all  three clusters. On the other hand, 
we note that a similar change in the internal uncertainty in $v_t$ 
 also increases  the internal spread detected in NGC~2808 by about a factor of 2. 
Hence, even if a real intrinsic spread in K abundance is present in NGC~104, NGC~6752, or
NGC~6809, it is significantly smaller than that detected in NGC~2808.
On the other hand, we must be aware that such very weak (but significant) 
correlations between abundance ratios can also be produced by  small systematics errors 
in the atmospheric parameters or by subtle inadequacies of the model atmosphere, 
displaying an effect smaller than the uncertainty on the individual abundance estimate.

In the cases of Mg and Al, where a lower number of stars with simultaneous 
measures of K  is available (Figs.~\ref{kmg} and \ref{kal}) no obvious correlation is found.
However, we note that, in the three GCs considered here, [K/Fe] and [Mg/Fe] exhibit the same
behaviour with respect to [Al/Fe], i.e., they are virtually constant while [Al/Fe] spans about one
full dex, while stars in NGC~2808 show a weak but significant positive correlation.

\begin{figure}
\includegraphics[width=84mm]{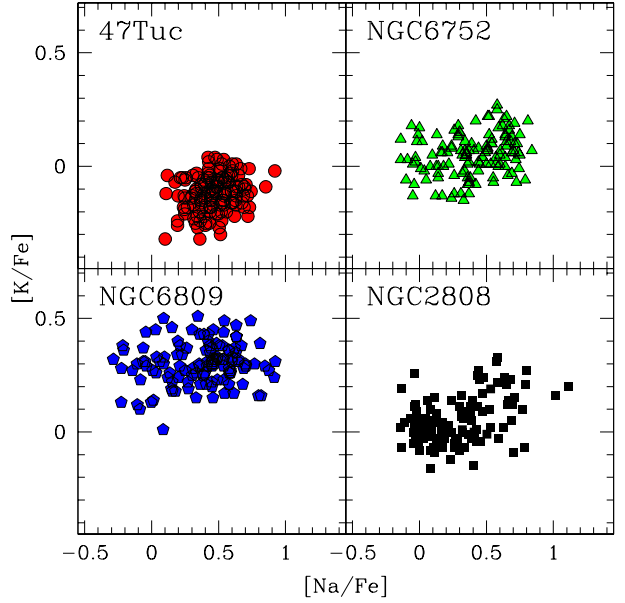}
\caption{Behaviour of [K/Fe] as a function of [Na/Fe].
Same symbols as in Fig.~\ref{vturb1}. Black squares represent the comparison stars 
of NGC~2808 \citep{m15}.}
\label{kna}
\end{figure}

\begin{figure}
\includegraphics[width=84mm]{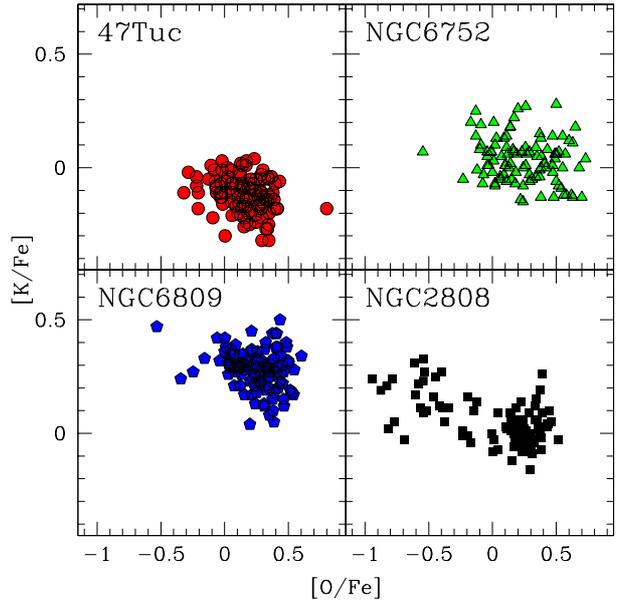}
\caption{Behaviour of [K/Fe] as a function of [O/Fe] (same symbols as in Fig.~\ref{kna}).}
\label{ko}
\end{figure}

\begin{figure}
\includegraphics[width=84mm]{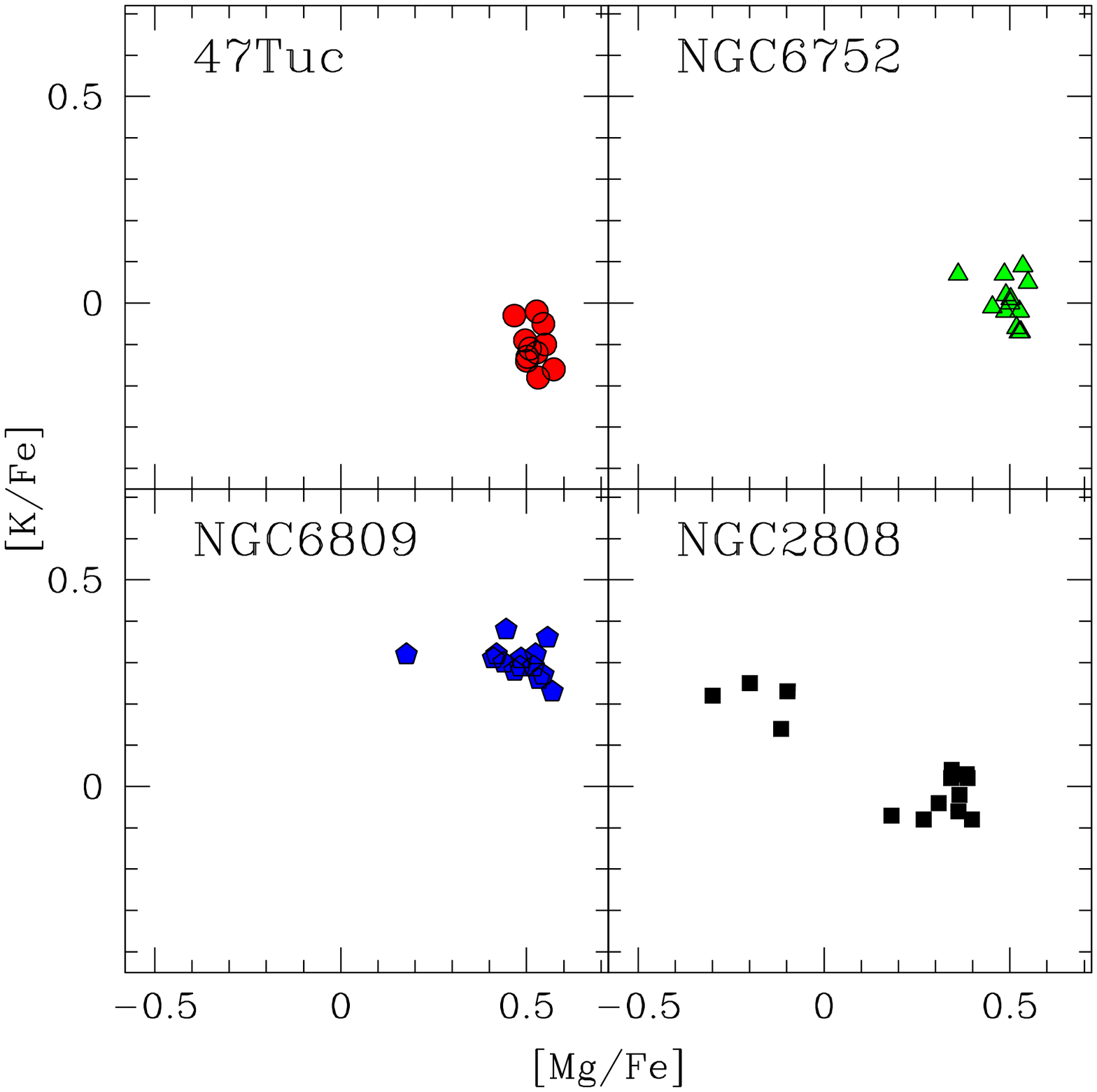}
\caption{Behaviour of [K/Fe] as a function of [Mg/Fe], only for the UVES targets 
(same symbols as in Fig.~\ref{kna}).}
\label{kmg}
\end{figure}

\begin{figure}
\includegraphics[width=84mm]{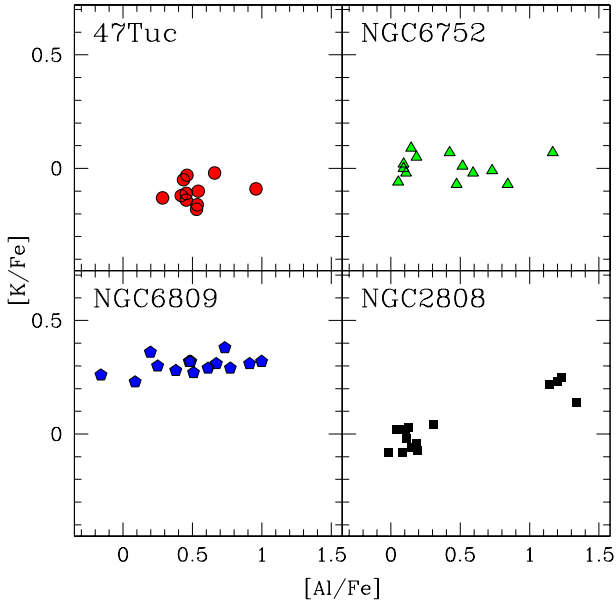}
\caption{Behaviour of [K/Fe] as a function of [Al/Fe], only for the UVES targets 
(same symbols as in Fig.~\ref{kna}).}
\label{kal}
\end{figure}


\section{Conclusions}

The main results of the K abundance analysis from high-resolution spectra in NGC~104, NGC~6752, and NGC~6809
can be summarised as follows:\\ 

{\sl (i)}~At variance with NGC~2419 and NGC~2808, where significant intrinsic star-to-star differences 
in the [K/Fe] abundance ratio have been detected, all  three GCs discussed here 
are formally compatible with a null [K/Fe] spread, which is in agreement with 
previous findings by \citet{carretta13}, based on smaller samples. \\

{\sl (ii)}~No hint of correlation between [K/Fe] and 
[Mg/Fe] or [Al/Fe] is found in NGC~104, NGC~6752, and NGC~6809; in particular none of them shows the [K/Fe]-[Mg/Fe] 
anti-correlation detected in NGC~2808 and NGC~2419. However the samples considered for
these correlations are small and there is no known Mg-deficient star in the considered clusters.\\

{\sl (iii)}~
[K/Fe] is found to correlate with [Na/Fe] and to anti-correlate to [O/Fe],
possibly hinting at the presence of a real, albeit very small, intrinsic spread in K abundances, which may
go undetected if the internal uncertainties are slightly overestimated. 
If this were the case, the spread should be significantly smaller than that measured in NGC~2808, typically
$\le 0.05$~dex. If these correlations reflect real physics relations (and not, e.g., a slight inadequacy 
of the measuring procedure),  
they provide further support to the hypothesis that  Potassium also has a role in the
multi-populations phenomenon. \\
 
All the evidence collected so far about K abundances in GCs, when considered within a self-enrichment scenario, 
suggests that K is an element produced within the same thermonuclear 
chains responsible of the enrichment/depletion of O, Na, Al, Mg etc.
However, the magnitude of the [K/Fe] spread, as well as the correlations with other abundance ratios, 
varies from cluster to cluster, emerging beyond our sensitivity limit only in clusters where the
process of GC self-enrichment reached the most extreme effects \citep{carretta14,m15}.
At this  stage this can be considered merely as a sound working hypothesis. The survey of K abundances must be extended, 
with particular attention to stars belonging to 
extreme second-generation stars \citep[as defined in][]{carretta14}, to get the observational basis sufficient 
to draw firmer conclusions.


\begin{table*}
\begin{minipage}{145mm}
\caption{Atmospheric parameters ($T_\mathrm{eff}$, logg and $v_t$), measured EWs of the K~I line at 7699 \AA\ , 
[K/Fe] abundance ratios and their uncertainties for all the target stars.}
\begin{tabular}{ccccccc}
\hline
Cluster &  ID &  $T_\mathrm{eff}$  &  log~g  &  $v_t$  &  EW       &  $[K/Fe]_\mathrm{NLTE}$  \\ 
        &     &    (K)      &         &  (km/s) &  (mA)  &   (dex)  \\ 
\hline
 NGC~104     &  1389  &  4568 &  2.09    &  1.65  &  184.50  &  -0.07$\pm$0.09\\  
 NGC~104     &  2608  &  3991 &  0.99    &  1.90  &  247.10  &  -0.11$\pm$0.10\\  
 NGC~104     &  2871  &  4609 &  2.17    &  1.63  &  174.80  &  -0.15$\pm$0.08\\  
 NGC~104     &  4373  &  4709 &  2.38    &  1.58  &  174.90  &  -0.06$\pm$0.08\\  
 NGC~104     &  5172  &  4560 &  2.08    &  1.65  &  183.30  &  -0.09$\pm$0.09\\  
 NGC~104     &  5270  &  3999 &  1.01    &  1.90  &  246.50  &  -0.10$\pm$0.10\\  
 NGC~104     &  5277  &  4237 &  1.48    &  1.79  &  219.50  &  -0.03$\pm$0.10\\  
 NGC~104     &  5640  &  4752 &  2.47    &  1.56  &  171.40  &  -0.05$\pm$0.06\\  
 NGC~104     &  6092  &  4627 &  2.21    &  1.62  &  182.60  &  -0.04$\pm$0.08\\  
 NGC~104     &  6808  &  4577 &  2.11    &  1.64  &  193.10  &  +0.04$\pm$0.08\\  
 NGC~104     &  7711  &  4649 &  2.25    &  1.61  &  186.00  &  +0.02$\pm$0.08\\  
 NGC~104     &  7904  &  4637 &  2.23    &  1.62  &  168.70  &  -0.20$\pm$0.08\\  
 NGC~104     &  9163  &  4418 &  1.81    &  1.71  &  198.40  &  -0.05$\pm$0.09\\  
 NGC~104     &  9268  &  4752 &  2.47    &  1.56  &  166.40  &  -0.11$\pm$0.08\\  
 NGC~104     &  9518  &  4463 &  1.90    &  1.69  &  194.70  &  -0.05$\pm$0.09\\   
 NGC~104     &  9717  &  4585 &  2.13    &  1.64  &  178.90  &  -0.12$\pm$0.09\\  
 \hline
\end{tabular}
\end{minipage}
\end{table*}

\begin{table*}
\begin{minipage}{145mm}
\caption{Average [K/Fe], observed ($\sigma_{obs}$) and intrinsic ($\sigma_\mathrm{int}$), number of measured stars and 
average [Fe/H] for the three target GCs.}
\begin{tabular}{cccccc}
\hline
Cluster &   $<[K/Fe]_\mathrm{NLTE}>$ & 
$\sigma_\mathrm{obs}$    & 
$\sigma_\mathrm{int}$    & 
$N_\mathrm{stars}$       & $<[Fe/H]>$ \\
  &    &   &  & &  \\
\hline
 NGC~104       &  --0.12$\pm$0.01  &  0.08  &  0.00$\pm$0.02     &  144  &  --0.76   \\  
 NGC~6752     &   +0.05$\pm$0.01  &  0.10  &  0.02$\pm$0.03      &  134  &  --1.55    \\   
 NGC~6809     &   +0.29$\pm$0.01  &  0.10  &  0.02$\pm$0.02      &  151  &  --1.93  \\  
 \hline
\end{tabular}
\end{minipage}
\end{table*}

\begin{table*}
\begin{minipage}{145mm}
\caption{Spearman correlation coefficient and associated two-tailed probability 
for the [K/Fe] abundance ratio against [O/Fe], [Na/Fe], [Mg/Fe], and [Al/Fe].}
\begin{tabular}{ccccc}
\hline
Cluster &   [K/Fe]--[O/Fe]& 
 [K/Fe]--[Na/Fe]    & 
 [K/Fe]-[Mg/Fe]    & 
 [K/Fe]-[Al/Fe]       \\
\hline
 NGC~104       &  $C_S$=--0.27  &  $C_S$=+0.20   & $C_S$=--0.30 &    $C_S$=+0.28    \\  
              &  P=0.0035      &  P=0.017       & P=0.37        &    P=0.40  \\ 
 NGC~6752     &  $C_S$=--0.18  &  $C_S$=+0.24   & $C_S$=--0.16  &    $C_S$=--0.06    \\  
              &  P=0.068      &  P=0.0065       & P=0.59        &    P=0.84  \\ 
 NGC~6809     &  $C_S$=--0.20  &  $C_S$=+0.11   & $C_S$=--0.37  &    $C_S$=+0.38    \\  
              &  P=0.035      &  P=0.22        & P=0.19         &    P=0.19  \\ 
 \hline
\end{tabular}
\end{minipage}
\end{table*}


\begin{acknowledgements}

We warmly thank the anonymous referee for suggestions that helped improve the paper.
We thank Eugenio Carretta for useful discussions throughout the course of this study.

\end{acknowledgements}


\end{document}